\documentclass[12pt]{article}

\usepackage{jcappub}

\usepackage{amsmath}
\usepackage{graphicx}
\usepackage{verbatim}
\usepackage{xfrac}
\usepackage{accents}

\newcommand{\be}{\begin{equation}}
\newcommand{\ee}{\end{equation}}

\newcommand{\ba}{\begin{eqnarray}}
\newcommand{\ea}{\end{eqnarray}}
\newcommand{\bs}{\begin{subequations}}
\newcommand{\es}{\end{subequations}}

\newcommand{\Ejf}{\overset{\sim}{\bf{E}}}

\title{Vacuum Cherenkov radiation and bremsstrahlung from disformal couplings}

\author[a]{Carsten van de Bruck,}
\author[b]{Clare Burrage}
\author[a]{and Jack Morrice}

\emailAdd{c.vandebruck@sheffield.ac.uk}
\emailAdd{Clare.Burrage@nottingham.ac.uk}
\emailAdd{app12jam@sheffield.ac.uk}

\affiliation[a]{Consortium for Fundamental Physics,\\
	School of Mathematics and Statistics,\\
	University of Sheffield,\\
	Hounsfield Road,\\
	Sheffield S3 7RH, United Kingdom.}
\affiliation[b]{School of Physics and Astronomy,\\ University of Nottingham,\\ Nottingham, NG7 2RD,\\
United Kingdom }

\abstract{The simplest way to modify gravity is to extend the gravitational sector to include an additional scalar degree of freedom. The most general metric that can be built in such a theory includes disformal terms, so that standard model fields move on a metric which is the sum of the space time metric and a tensor constructed from first derivatives of the scalar. In such a theory gravitational waves and photons can propagate at different speeds, and these can in turn be different from the maximum speed limit for matter particles. In this work we show that disformal couplings can cause charged particles to emit Cherenkov radiation and bremsstrahlung apparently in vacuum, depending on the background evolution of the scalar field.  We discuss the implications of this for observations of cosmic rays, and the constraints that arise for models of dark energy with disformal couplings. }

\begin{document}
\maketitle
\flushbottom

%
%
\section{Introduction\label{sec:intr}} 
The theory of General Relativity (GR) and its surrounding paradigm are unmatched for predictive success. Not even quantum field theory can boast agreement with experiment over such a vast range of scales \cite{Baker:2014zba}. Perhaps a victim of its own success, theorists have grown more and more focused on its shortfalls: the theory is non-renormalizable, divergent in the ultra violet regime and, when applied to cosmology, makes the uncomfortable prediction that 95\% of the matter in the universe is exotic, dark and intractable \cite{Copeland:2006wr}. Proposed modifications and upgrades abound. A popular set of such modifications are models of gravity in which two fields, not one, mitigate the force, and of this popular set, a subset are the disformal theories in which the gravitational geometry, $\hat{g}_{\mu\nu}$ and the matter geometry, $\tilde{g}_{\mu\nu}$ are related via the \emph{disformal} transformations 
\be
\tilde{g}_{\mu\nu} = \hat{g}_{\mu\nu} + D(\phi)\phi,_{\mu}\phi,_{\nu}
\ee
for some additional gravitational scalar field $\phi$ \cite{Bekenstein:1992pj}\footnote{We have not written the most general transformation here. There could be a conformal part in front of $\hat{g}_{\mu\nu}$ as well, but we are dealing with electromagnetism in this paper so that that term is irrelevant.}.

In the history of disformal terms in gravity theories there exists a multiplicity of purpose. They were geometry corrections to GR in compactifications of higher dimensional brane-world gravity theories, but they were also utilized in the literature to vary the relationship between the speed of light and the speed of gravitational waves, which could solve the horizon and flatness problems of early universe cosmology without recourse to a potential-driven inflationary phase \cite{Clayton:2001rt}\cite{Kaloper:2003yf}\cite{Magueijo:2008sx}\cite{Magueijo:2010zc}\cite{vandeBruck:2015tna}. (Such theories are now very tightly constrained by observations \cite{Magueijo:2003gj}.) This second aspect of disformal theories -- their tendency to distort the light cones of fundamental fields with respect to each other -- is what concerns us in this work, however here we focus on the late, rather than inflationary, universe.

Coupling universally to all matter has been constrained quite severely via global tests in cosmology \cite{Brax:2014vla}, or local tests in the solar system \cite{Ip:2015qsa} or the laboratory \cite{Brax:2015hma,Brax:2016did}, which has led some to postulate that disformal couplings can, for example, only be between the scalar and dark matter \cite{Zumalacarregui:2012us}\cite{vandeBruck:2015ida}. As the nature of dark matter is poorly understood, the constraints of disformal couplings to it are rather weak. This idea of species selectivity opens the door, though, to varying interaction strengths with respect to varying types of matter (dark, baryonic,  electromagnetic sector, etc.); if strengths can vary from species to species, there is little theoretical motivation to assume that the coupling to the standard model particles is negligibly small. Relaxing this assumption will invariably lead to observable deviations from standard matter theory. 

A handful of these deviations must be in the form of novel radiation processes. Due to the variation in the relative speeds of photons and gravitons in disformal theories, it remains an open question as to whether charged particles, disformally coupled, can Cherenkov radiate in vacuum. In this work we unambiguously demonstrate that this is indeed the case, and deduce the  conditions that must be met in order for this to occur. We will also discover that another radiative interaction channel will open under those same model conditions, a channel that depends on the dynamics of the theory's speed of light. For reasons that will become clear, we dub this interaction \emph{vacuum bremsstrahlung}. 

In \cite{vandeBruck:2013yxa} it was shown that in order to induce spectral distortions in the CMB via gravity modifications, a necessary ingredient was that the geometry of space-time experienced by photons and that of the rest of the Standard Model must vary disformally with respect to each other. Hence, we consider the following action
\be\label{eq:action1}
\mathcal{S} 
= \mathcal{S}_{\mathrm{grav}}[\hat{g}_{\mu\nu},\phi]
+\mathcal{S}_{\mathrm{matter}}[\tilde{g}^{(m)}_{\mu\nu}]
+ \mathcal{S}_{\mathrm{EM}}[\tilde{g}^{(r)}_{\mu\nu},A^{\mu}]
+ \mathcal{S}_{\mathrm{int}}~,
\ee
where the definition of the interaction terms will be clarified in the next section, and the metrics relate in the following way
\bs
\ba
\tilde{g}^{(m)}_{\mu\nu} &=& \hat{g}_{\mu\nu} + D^{(m)}(\phi)\phi,_{\mu}\phi,_{\nu} \\
\tilde{g}^{(r)}_{\mu\nu} &=& \hat{g}_{\mu\nu} + D^{(r)}(\phi)\phi,_{\mu}\phi,_{\nu}.
\ea
\es
We refer to $\tilde{g}^{(m)}$ as the \emph{matter metric}, $\tilde{g}^{(r)}$ as the \emph{electromagnetic metric}, and $\hat{g}$ the \emph{gravitational metric}.

In the next section we refine this action and restrict our attention to a minimal subsystem in which to cleanly explore novel radiative processes, but in the meantime this schematic action, Eq. \eqref{eq:action1}, highlights a key point: there are three metrics in our theory, all related by disformal transformations, so there are three different frames within which to make calculations, and three representations of each field. In standard scalar-tensor theory, it is commonplace to perform computations in the Einstein frame, where the gravitational action is of Einstein-Hilbert form (quantities are defined with respect to $\hat{g}$), and reserve physical interpretation for the Jordan frame (everything expressed in terms of $\tilde{g}^{(m)}$), however, for this work, we will find that while physical interpretation is easiest in the Jordan frame, it is in fact the \emph{Electromagnetic frame} (expressing the full action in terms of $\tilde{g}^{(r)}$) in which calculations are simplest. This will hopefully become clear as we unveil the calculation.

As we have mentioned above, we find that two radiation channels are open to a disformally coupled charged particle, provided certain radiation conditions are satisfied: vacuum Cherenkov and bremsstrahlung radiation. Both of which we consider in what follows. In section \ref{sec:cher} we introduce the gravitational part of the action \eqref{eq:action1}, and specify a small charged particle-and-field subsystem -- adequate to determine the conditions under which vacuum Cherenkov radiation occurs in a disformal theory. 
In this section we present Maxwell's equations with disformal couplings present, then solutions and finally constraints on model parameters from collider based experiments. In section \ref{sec:brem} we present the case for bremsstrahlung, define the relevant parts of the action, derive equations of motion, and discuss the conditions to be met for its presence. We do not use vacuum bremsstrahlung to place theory constraints in this paper, but simply offer an illustration as to the scale of the effect in a cosmology setting using cosmic rays. Our conclusions can be found in section \ref{sec:conc}.

%
%
\section{Vacuum Cherenkov radiation\label{sec:cher}}

\subsection{Action\label{sec:action}}
The salient feature of our model is the disformal coupling to radiation; we ask what novel changes this detail will introduce into the theory of electromagnetism. The electromagnetic sector is specified by the terms $\mathcal{S}_{\mathrm{field}} + \mathcal{S}_{\mathrm{interaction}}$ which we write as
\bs\label{eq:action2}
\begin{equation}
{\cal S}_{\rm field} =
-\frac{1}{4\mu_0} \int d^4 x \sqrt{-g^{(r)}} g_{(r)}^{\mu\nu} g_{(r)}^{\alpha\beta} F_{\mu\alpha}F_{\nu\beta}
\end{equation}
and
\begin{equation}
{\cal S}_{\rm interaction} = 
\int d^4 x \sqrt{-g^{(m)}} j^\mu A_\mu,
\end{equation}
\es
where $j^\mu$ is a four--current, describing the motion of a charged particle. Note that gauge invariance implies charge conservation, i.e. we have $\nabla_\mu j^\mu = 0$, where the covariant derivative is with respect to the metric $g_{\mu\nu}^{(m)}$. As it will be useful for the subsequent calculations, we will write the action in terms of the matter metric $g^{(m)}_{\mu\nu}$. Note that 
\begin{equation}\label{eq:trans}
g^{(r)}_{\mu\nu} 
= g_{\mu\nu}^{(m)} + \left( D^{(r)} - D^{(m)}  \right)\phi_{,\mu}\phi_{,\nu} 
:= g_{\mu\nu} + B\phi_{,\mu}\phi_{,\nu} ,
\end{equation}
where in the last equation we have dropped the tilde and written $g_{\mu\nu} := g_{\mu\nu}^{(m)}$ to simplify notation. We emphasise that $B$ measures the difference between the disformal couplings $D^{(r)}$ and $D^{(m)}$. Then, in terms of this metric the electromagnetic sector becomes
\begin{equation}\label{eq:field_action}
{\cal S}_{\rm field} = -\frac{1}{4\mu_0} \int d^4 x \sqrt{-g}Z\left[g^{\mu\nu}g^{\alpha\beta} - 2 \gamma^2 g^{\mu\nu}\phi^{,\alpha}\phi^{,\beta} \right]F_{\mu\alpha}F_{\nu\beta}~,
\end{equation}
with
\begin{equation}\label{def:Z}
Z := \sqrt{\frac{g^{(r)}}{g}} = \sqrt{ 1 + Bg^{\mu\nu}\partial_\mu \phi \partial_\nu \phi  }
\end{equation}
and
\begin{equation}
\gamma^2 = \frac{B}{1+Bg^{\mu\nu}\partial_\mu \phi \partial_\nu \phi}~.
\end{equation}
Note that the dynamics of $\phi$ are not specified at this point; it is a generic scalar field. We also have not specified the gravitational sector at this point. The work below holds for generic modified gravity theories. Only later we will be specific when we discuss constraints on the theory. 

%
\subsection{Disformal Maxwell's equations}
The electromagnetic field equation can be readily obtained from this action: 
\ba\label{generalfield}
\nabla_\epsilon\left(Z F^{\epsilon\rho} \right) - \nabla_\epsilon \left(  Z\gamma^2 \phi^{,\beta}\left( g^{\epsilon\nu}\phi^{,\rho} -  g^{\rho\nu}\phi^{,\epsilon} \right)F_{\nu\beta} \right) = - \mu_0j^\rho~.
\ea
From now on, we will consider the case of flat space, i.e. $g_{\mu\nu} = \eta_{\mu\nu}$ (we remind the reader that matter moves on geodesics with respect to this metric) and write $A^\mu = (\Phi/c,{\bf A)}$ and $j^\mu  = (c\rho,{\bf j})$. Then, working in the disformal Lorenz gauge $\nabla\cdot{\bf A} = - \dot{\Phi}/(cZ)^2$, the dot denoting the derivative with respect to time $t$, Eq. \eqref{generalfield} becomes 
\bs
\begin{eqnarray}
\left( \nabla^2 - \frac{1}{c^2Z^2}\frac{\partial^2}{\partial t^2} \right) \Phi 
&=& - \frac{Z}{\epsilon_0} \rho \label{Phieq0} \\
\left( \nabla^2 - \frac{1}{c^2Z^2}\frac{\partial^2}{\partial t^2} \right) {\bf A} 
&+& \frac{1}{c^2}\frac{\dot Z}{Z}\left( \nabla \dot\Phi + \dot {\bf A} \right) 
= - \frac{\mu_0}{Z} {\bf j} \label{Aeq0}~.
\end{eqnarray}
\es
In deriving the last equation, we made use of the  identity $\nabla (\nabla\cdot {\bf V}) = \nabla^2 {\bf V} + \nabla \times (\nabla \times {\bf V})$ and defined in the usual way $\epsilon_0 = 1/\mu_0 c^2$. For the case that the scalar is time dependent only, Maxwell's equations read
\bs\label{eq:max}
\begin{eqnarray}
\nabla \cdot {\bf E} &=& \frac{Z}{\epsilon_0} \rho \\
\nabla \times {\bf B} &=& \frac{\mu_0}{Z}{\bf j} + \frac{\mu_0}{Z}\frac{\partial}{\partial t} \left( \frac{\epsilon_0}{Z}{\bf E} \right)\\
\nabla \cdot {\bf B} &=& 0 \\
\nabla \times {\bf E} + \frac{\partial {\bf B}}{\partial t} &=& 0~,
\end{eqnarray}
\es
where ${\bf E}$ and ${\bf B}$ are defined in the usual way: 
\begin{eqnarray}
{\bf E} = -\nabla \Phi - \frac{\partial {\bf A}}{\partial t}~~~{\rm and}~~~ {\bf B} = \nabla\times {\bf A} ~.
\end{eqnarray}
Momentarily considering a vacuum (i.e. $\rho = 0$, ${\bf j}= {\bf 0}$), and assuming that $Z$ is constant, from Maxwell's equations we can derive the following wave equations for the fields ${\bf E}$ and ${\bf B}$
\bs
\begin{eqnarray}
-\frac{1}{c^2Z^2} \frac{\partial^2 {\bf E}}{\partial t^2} + \nabla^2 {\bf E} &=& 0~ \\
-\frac{1}{c^2Z^2} \frac{\partial^2 {\bf B}}{\partial t^2} + \nabla^2 {\bf B} &=& 0~,
\end{eqnarray}
\es
which shows that, in the absence of charges and with $Z$ constant, electromagnetic fields propagate with a modified speed\footnote{It was shown in \cite{vandeBruck:2015rma} that the fine-structure coupling `constant' is not constant in this theory.}. Prompted by this observation, we define more generally
\begin{equation}\label{sol}
c_s(t) := c Z(t)  = \left( c^2 - B\dot\phi^2  \right)^{1/2}. 
\end{equation} 
%

The set of equations \eqref{eq:max} quite clearly suggest that we can go further; an effective speed of light here arises as a consequence of the fact that the disformal couplings modify spacetime geometry and hence distort the electromagnetic vacuum, producing an \emph{effective medium} for the electromagnetic field, whose permeability, $\mu_0$, and permittivity, $\epsilon_0$, of free space are modified by the scalar interaction. We thus also make the definitions
\be
\mu(t) := \frac{\mu_0}{Z(t)} ~~~{\rm and}~~~ \epsilon(t) := \frac{\epsilon_0}{Z(t)}
\ee
to physically characterize this new effective vacuum, and, subsequently, the auxiliary fields 
\be
{\bf H}:=\frac{1}{\mu(t)} {\bf B} ~~~{\rm and}~~~ {\bf D}:=\epsilon(t) {\bf E}.
\ee

Given this effective medium formulation, we can now ask how the energy density will change in the field due to time evolution of our scalar field. In terms of the auxiliary fields the first two Maxwell equations simplify as follows: 
\bs
\ba
\nabla\cdot{\bf D} = \rho \\
\nabla\times{\bf H} - \bf{\dot{D}} = {\bf j},
\ea
\es
from which we obtain Poynting's theorem in our theory: 
\be\label{eq:poynt}
\frac{d}{dt}(U_E + U_H) = \frac{\dot{Z}}{Z}( U_E + U_H ) - {\bf E}\cdot{\bf j} 
-\nabla\cdot(\underbrace{{\bf E}\times{\bf H}}_{{\bf S}})~.
\ee
Here we have defined the field energy densities 
\be
U_E:=\frac{1}{2}\epsilon(t) |{\bf E}|^2, ~~~~~ U_H:=\frac{1}{2}\mu(t)|{\bf H}|^2~,
\ee
and identified the standard Poynting vector $\bf{S} = \bf{E}\times\bf{H}$, which we will use to compute the energy lost by a charged particle in superluminal flight in the next section.  

To summarize, we have found that when the scalar is time dependent only, our field theory with disformal couplings reduces to that of an electromagnetic field in an effective linear medium, whose resistance to the formation and evolution of field disturbances ($\epsilon$, $\mu$) will depend on $Z(t)$: the ratio of the two metric determinants. This establishes an interesting conceptual link between the geometry of space and the physical response of the fields defined on it. The link should strengthen the reader's intuition that many analogues of electricity in linear media should carry through to this model.

%
\subsection{Field solutions and the Cherenkov radiation condition}
As a first application of the model, we will investigate under which circumstances Cherenkov radiation can occur. We follow the calculation in \cite{lecturenotes}. The speed of light $c_s$, given by Eq. (\ref{sol}), is smaller than the bare speed of light $c$ if the field evolves in time, i.e. if $\dot\phi$ is non-vanishing. A charged particle can then move faster than $c_s$ and this is the situation which we will now study. Let us therefore consider a moving particle with charge $q$, for which
\bs
\ba
\rho({\bf x},t) &=& q \delta({\bf x} - {\bf x}_p(t))\\
{\bf j}({\bf x}, t) &=& \rho {\bf v}~,
\ea
\es
with ${\bf x}_p(t)$ the time dependent position in 3-space of the moving particle, and ${\bf v} = \dot{{\bf x}}_p$ the velocity. Furthermore, we assume in this section that $\phi = \phi(t)$ with $c_s=cZ={\rm constant}$.

Then, considering the Fourier space components one obtains from Eq.(\ref{Phieq0}) 
\be
\Phi_{k} = \frac{2\pi q }{\epsilon} \frac{\delta(\omega - {\bf k}\cdot {\bf v})}{k^2 - \dfrac{\omega^2}{c_s^2}},
\ee
and Eq.(\ref{Aeq0}) can be solved to find 
\begin{equation}\label{A_k}
{\bf A}_k = 2 \pi q\mu \frac{\delta(\omega - {\bf k}\cdot{\bf v})}{k^2 - \dfrac{\omega^2}{c_s^2}}{\bf v}~.
\end{equation}
As a consistency check, these solutions imply the Lorenz--gauge condition ${\bf k}\cdot{\bf A}_k = \omega \Phi_k/c_s^2$. 
The Fourier coefficients of ${\bf B}$ are related to ${\bf A}_k$ via ${\bf B}_k = i{\bf k}\times {\bf A}_k$ and the Fourier coefficients of ${\bf E}$ are given by ${\bf E}_k = -i{\bf k} \Phi_k + i \omega {\bf A}_k$. We find
\begin{equation}
{\bf B_k}({\bf k},\omega) = 2\pi i q \mu \frac{{\bf k}\times{\bf v}}{k^2 - \dfrac{\omega^2}{c_s^2}}\delta(\omega - {\bf k}\cdot{\bf v})
\end{equation}
and 
\begin{equation}
{\bf E}_k({\bf k},\omega) = - \frac{2\pi iq}{\epsilon} \frac{{\bf k} - \dfrac{\omega}{c_s^2}{\bf v}}{k^2 - \dfrac{\omega^2}{c_s^2}}\delta(\omega - {\bf k}\cdot{\bf v})~.
\end{equation}
To find the energy loss along the particle's trajectory, we assume without the loss of generality that the particle moves along the $z$-axis with velocity ${\bf v} = (0,0,v)$, and that the observer is located at a distance $r$ from the $z$-axis. The energy loss per unit length is then given by the integral 
\begin{equation}\label{energyloss}
- \frac{d{\cal E}}{dz} = -2\pi r \int E_z({\bf r},t) B_{\phi}({\bf r},t)dt = -r \int E_z({\bf r},\omega) H^*_{\phi}({\bf r},\omega)d\omega ,
\end{equation}
where 
\begin{eqnarray}
E_z({\bf r},\omega) &=& \frac{1}{(2\pi)^3} \int d^3 k E_z ({\bf k},\omega) e^{i{\bf k}\cdot {\bf r}} {~~~~{\rm and}} \nonumber \\
H_\phi({\bf r},\omega) &=& \frac{1}{(2\pi)^3} \int d^3 k H_\phi ({\bf k},\omega) e^{i{\bf k}\cdot {\bf r}}.
\end{eqnarray}
Evaluating the integrals for $\beta = v/c_s > 1$, we find 
\begin{eqnarray}
E_z({\bf r},\omega) &=& 
\frac{iq\mu\omega}{2\pi}\left[ 1 - \frac{1}{\beta^2}  \right] e^{i\omega z/\beta c_s} K_0(\alpha r), \\
H_\phi({\bf r}, \omega) &=& 
\frac{\alpha q}{2\pi}e^{iz\omega/\beta c_s} K_{1}\left(\alpha r  \right)~,
\end{eqnarray}
where $\alpha = -(i\omega/c_s)\sqrt{1-\beta^{-2}}$. Note that for large $\alpha r$, $K_0 (\alpha r) \approx \sqrt{\pi/(2\alpha r)} \exp(-\alpha r)$, so these represent outgoing waves if $\beta>1$. The expressions for $E_z$ and $H_\phi$ are identical to those for electromagnetic waves propagating through a medium, leading to Cherenkov radiation for $v>c_s$. The integral (\ref{energyloss}) can be evaluated for $|\alpha | r \gg1$, giving 
\begin{equation}
- \frac{d{\cal E}}{dz} = \frac{1}{4\pi \epsilon_0}\frac{e^2}{c^2} \int \omega \left( 1 - \frac{1}{\beta^2} \right) d\omega~.
\end{equation}

%
\subsection{Constraints\label{sec:constraints}}
We will now discuss constraints on the model. So far, the scalar field has been completely unspecified. The only assumption we have made is that it is disformally coupled to the electromagnetic sector. To specify the dynamics of the field, we have to specify the action for it, and in the following we assume that the gravitational sector is of standard Einstein form, together with a canonical scalar field. The form of $\mathcal{S}_{grav}$ in equation \eqref{eq:action1} we then chose to be
\be\label{eq:grav_action}
\mathcal{S}_{\mathrm{grav}}
=\int d^4x\sqrt{-g}\left[ \frac{R(g)}{2\kappa} - \frac{1}{2}g^{\mu\nu}\phi,_{\mu}\phi,_{\nu} - V(\phi)  \right], 
\ee
where we assume that $\phi$ is the scalar field responsible for the accelerated expansion of the universe at late times and we assume $\hat{g}_{\mu\nu} = g_{\mu\nu}$, which implies that we set $D^{(m)}=0$.

There are direct constraints on isotropic deviations of the speed of light from unity  from laboratory experiments \cite{Michimura:2013kca, Baynes:2012zz} at the level of $|1- c_s/c|< 10^{-10}$, however stronger constraints arise from searches  for Cherenkov radiation from particles in vacuum.  These can be done in terrestrial experiments, with bounds $|1- c_s/c|< 10^{-11}$ coming from the absence of vacuum Cherenkov radiation from $104.5 \mbox{ GeV}$ electrons and positrons at LEP \cite{Hohensee:2009zk}. Indeed, the energetics of the LEP beam were so well understood that measurements of the synchrotron emission rate indicate that any deviation of the speed of photons is constrained by $|1-c_s/c|< 5 \times 10^{-15}$, \cite{Altschul:2009xh}.  Observations of high energy cosmic rays provide significantly tighter constraints; the lack of vacuum Cherenkov radiation from high energy electrons and neutrinos propagating over astronomical distances constrains $|1-c_s/c|< 10^{-20}$ \cite{Stecker:2014xja, Diaz:2013wia, Stecker:2013jfa}, however these constraints come with some uncertainty about the high energy dynamics of the source of the cosmic ray. 

To translate these constraints into constraints on disformal electrodynamics, we assume now that the scalar field is slowly evolving and plays the role of dark energy. Firstly, we assume the constraint $|1-c_s/c|<5\times 10^{-15}$. The speed of light $c_s$ should not deviate drastically from one, so we can expand $Z \approx 1 - B\dot\phi^2/2c^2$. 

The Friedmann equation evaluated today reads
\be
3H^2_0 = \kappa \left( \rho c^4 + \frac{1}{2}\dot{\phi}^2 + c^2V \right)
\ee
for the bare speed $c$. If we assume that the scalar $\phi$ plays the role of dark energy then we have 
\be\label{eq:OmegaDE}
\Omega_{\rm DE} 
= \frac{\kappa}{6}\left(\frac{\dot{\phi}}{H_0}\right)^2+\frac{\kappa c^2V}{3H_0^2} \simeq 0.7,
\ee
where $\Omega_{\rm DE}$ is the dark energy density parameter. The equation of state of dark energy is
\begin{equation}
w_{\rm DE,0} = \frac{\dot\phi^2 - 2c^2V}{\dot\phi^2 + 2c^2V}
\end{equation}
which, combined with equation \eqref{eq:OmegaDE} gives 
\begin{equation}
\frac{\kappa\dot\phi^2}{2c^2} =  \frac{3}{2}\Omega_{\rm DE}H_0^2(1+\omega_{\rm DE,0}).
\end{equation}
Hence, the constraint can be written as $B\dot\phi^2/2c^2 < 5\times 10^{-15}$ or, expressed as a dimensionless ratio:
\begin{equation}\label{eq:constraintonM}
\frac{B_0 H_0^2}{\kappa} < \frac{10^{-14}}{3\Omega_{\rm DE}(1+\omega_{\rm DE,0})}.
\end{equation}

\begin{figure}\label{fig:cherenkov}
\begin{center}
\includegraphics[width=0.7\textwidth]{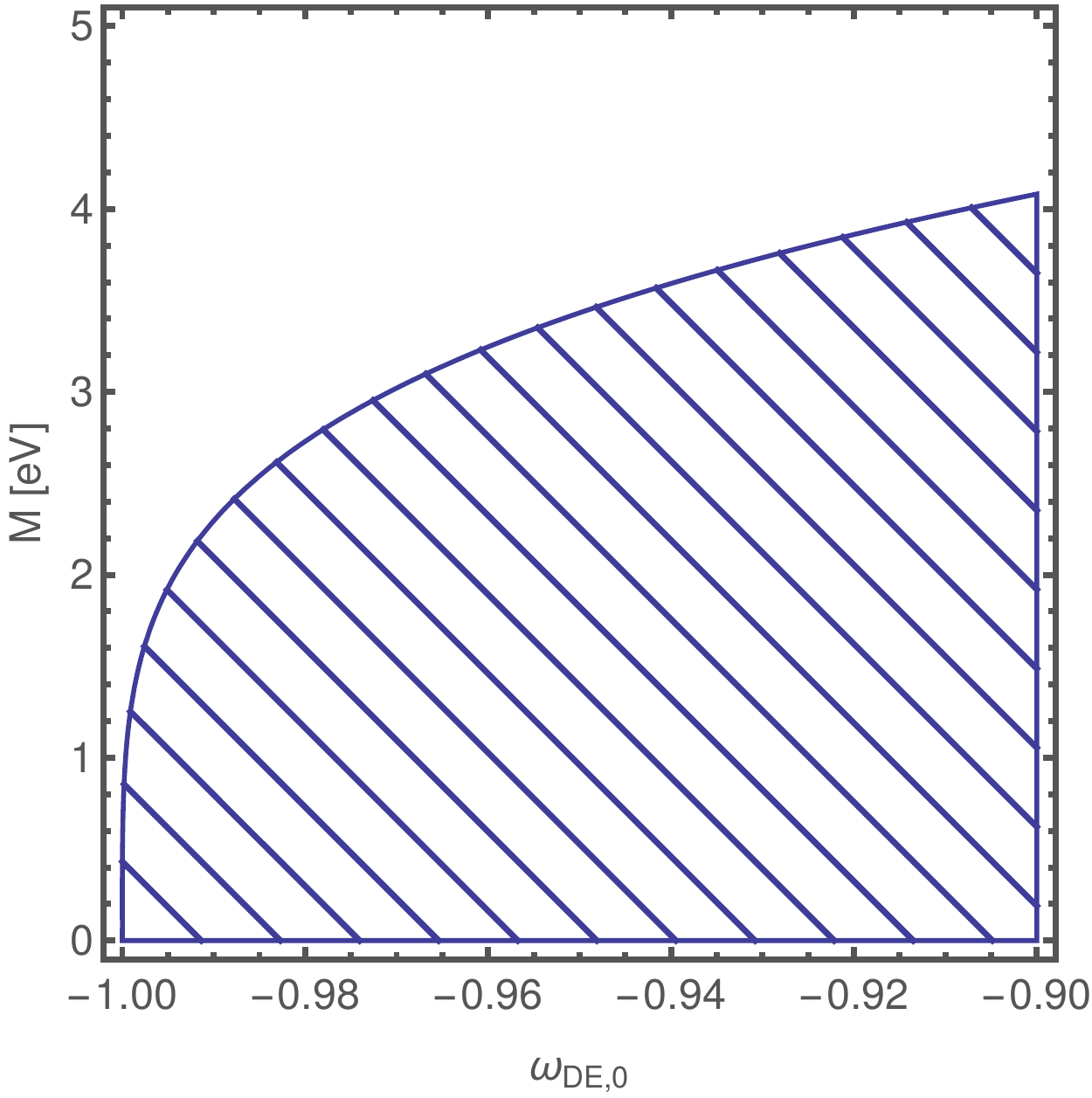}
\caption{Cherenkov radiation in vacuum constraints the energy scale $M$, defined in Eq. (\ref{eq:M}), as a function of the current dark energy equation of state $\omega_{\rm DE,0}$. The shaded region is ruled out by bounds coming from the LEP constraint $|1-c_s/c|< 5 \times 10^{-15}$. As the dark energy equation of state approaches $-1$, $\dot\phi$ approaches 0 and hence $c_s \rightarrow c$ and the constraint on $M$ vanishes in this limit.}
\end{center}
\end{figure}

In Fig. \ref{fig:cherenkov} we show the constraint on the energy scale:
\be\label{eq:M}
M := \left(\frac{c \hbar^3}{B_0}\right)^{1/4}
\ee
as a function of the dark energy equation of state $\omega_{\rm DE,0}$, measured today, setting $\Omega_{\rm DE} = 0.7$. We remind the reader that constraints of this type will always place limits on the difference between the disformal couplings to matter and radiation, since $B = D^{(r)} - D^{(m)}$ (see eq. (\ref{eq:trans})), though we have set $D^{(m)}=0$ here. 

%
%
\section{Vacuum Bremsstrahlung\label{sec:brem}}
We have seen that the particle will emit Cherenkov radiation in vacuum, if the effective speed of light $c_s$ drops below the particle speed $v$. A natural question to ask, given the close resemblance at the classical level our model has with that of a linear dielectric medium, is whether or not other radiative channels are open in the presence of a disformal coupling. In this section we derive the conditions for vacuum bremsstrahlung.

We are particularly interested in the possibility that charged cosmic rays might emit bremsstrahlung due to the evolution of the scalar $\phi$ in the cosmological background. Therefore we generalize our calculations to an expanding background with $c_s$ now time dependent in what follows. 

\subsection{Action\label{sec:action_brem}}

We consider again a subsystem of the action in \eqref{eq:action1} where a single charged particle in flight couples to an electromagnetic field: $\mathcal{S}_{\mathrm{field}}+\mathcal{S}_{\mathrm{int}}$, (see equation \eqref{eq:action2}), however,  we now work on an expanding background, and so we chose comoving coordinates such that
\be
g_{\mu\nu} = a^2(\tau)\eta_{\mu\nu},
\ee 
where $\tau$ is the conformal time, related to the physical time by $dt = ad\tau$, hence
\be
g^{(r)}_{\mu\nu} = a^2(\tau)\left(\eta_{\mu\nu} + \frac{B}{a^2}\phi_{,\mu}\phi_{,\nu}\right)
:= a^2 h^{(r)}_{\mu\nu}.
\ee
The gravitational action is still given by Eq. \eqref{eq:grav_action} and we assume that the scalar field $\phi$ depends on time only. 
Then, as $\mathcal{S}_{\mathrm{field}}$ is conformally invariant, we have 
\begin{equation}
{\cal S}_{\rm field} =
-\frac{1}{4\mu_0} \int d^4 x  ~Z h_{(r)}^{\mu\nu} h_{(r)}^{\alpha\beta} F_{\mu\alpha}F_{\nu\beta},
\end{equation}
where, recalling the definition of $Z$ (Eq. \eqref{def:Z}), we have now 
\be
Z=\sqrt{-h^{(r)}} = \sqrt{1+\frac{B}{a^2}\eta^{\mu\nu}\phi,_{\mu}\phi,_{\nu}}~.
\ee
For the interaction term, we define the \emph{comoving current} 
\be
J^{\mu} := \sqrt{-g} j^{\mu},
\ee
so that 
\begin{equation}
{\cal S}_{\rm int} = 
\int d^4 x J^\mu A_\mu.
\end{equation}
As $\nabla_{\mu}j^{\mu}=0$ (see section \ref{sec:action}), we have that the comoving current is conserved with respect to the flat metric $\eta_{\mu\nu}$, i.e. 
\be
\partial_{\mu}J^{\mu} 
= \partial_{\mu}\left( \sqrt{-g} j^{\mu} \right)
= \sqrt{-g} \nabla_{\mu}j^{\mu} = 0,
\ee
where we have used that $\sqrt{-g}\nabla_{\mu}v^{\mu} = \partial_{\mu}\left( \sqrt{-g} v^{\mu} \right)$ for any 4 vector $v^{\mu}$ and metric $g$. Lastly, we consider a point-like charged particle whose motion can be described by a curve ${\bf x}_p(\tau)$, and, since $\partial_{\mu}J^{\mu}=0$, we can define $J^{\mu} = (c\Omega,{\bf J})$ such that
\bs
\ba
\Omega({\bf x}, \tau) &:=& Q\delta({\bf x} - {\bf x}_p(\tau))\\
{\bf J}({\bf x}, \tau) &:=& \Omega{\bf V}
\ea
\es
for ${\bf V} := d{\bf x}_p / d\tau$ and $Q$ the charge of the particle. By construction this ansatz satisfies the continuity equation. Comparing this to the physical current, expressed in terms of the physical time, $t$, it is straightforward to show that $j^{\mu\prime} = (c\Omega/a^3, {\bf v} \Omega/a^3)$, and hence the charge density dilutes as $a^{-3}$, as it must in isotropically expanding space. It is also clear that, for $a(\tau)$ an arbitrary function, light still propagates with velocity 
\be
c_s(\tau) = Z(\tau)c.
\ee

\subsection{Disformal Maxwell's equations in expanding space}
The electromagnetic field equations can be readily obtained from the action specified in section \ref{sec:action_brem} as before; they are the expanding-space counterpart to Eq.s \eqref{eq:max}:
\bs\label{eq:max_brem}
\begin{eqnarray}
\nabla \cdot {\bf E} &=& \frac{Z}{\epsilon_0} \Omega~, \\
\nabla \times {\bf B} &=& \frac{\mu_0}{Z}{\bf J} + \frac{\mu_0}{Z}\frac{\partial}{\partial \tau} \left( \frac{\epsilon_0}{Z}{\bf E} \right)~,\\
\nabla \cdot {\bf B} &=& 0~, \\
\nabla \times {\bf E} + \frac{\partial {\bf B}}{\partial \tau} &=& 0~.
\end{eqnarray}
\es
In these equations, $\nabla$ is the \emph{flat} 3-space derivative operator. Even though space is expanding, this is valid, as the dependance of the system on the scale factor $a$ was absorbed by the field redefinitions in the previous section. 

From definition \eqref{sol} we see that if the speed of light were to vary in time in some coordinate system with time $t$, there would naturally exist some new system of coordinates such that this speed remains constant. If the particle were to travel with fixed velocity in the original system, it would appear to accelerate with respect to these new coordinates in which $c_s$ is constant. The electric field thus `sees' an accelerating charge. We would expect such a field to radiate accordingly, and indeed this is what we will find.

To make this intuition mathematically precise, we must consider a case more general than the previous sections, whereby $Z(t)$ becomes now an arbitrary function of time. Some suitable field and coordinate redefinitions will help us find solutions in this new case. Considering again the disformal Maxwell's equations, \eqref{eq:max_brem}, the following redefinitions are useful:
\be\label{eq:redefs}
\Ejf:=\frac{\bf E}{Z(\tau)}, \quad \tilde{\bf J}:=\frac{\bf J}{Z(\tau)}, \quad d\tilde{\tau} := Z(\tau)d\tau.
\ee
These fields obey the following equations:  
\bs\label{eq:maxJF}
\begin{eqnarray}
\nabla \cdot \Ejf &=& \frac{\Omega}{\epsilon_0}~, \\
\nabla \times {\bf B} &=& \mu_0\tilde{\bf J} + \mu_0\epsilon_0\frac{\partial}{\partial {\tilde \tau}} \left( \Ejf \right)~,\\
\nabla \cdot {\bf B} &=& 0~, \\
\nabla \times \Ejf + \frac{\partial {\bf B}}{\partial \tilde{\tau}} &=& 0~.
\end{eqnarray}
\es
%
This set of equations allows us make the standard gauge field definitions: ${\bf B} = \nabla\times{\bf A}$ as before, and now 
\be\label{eq:Epot}
\Ejf = -\nabla \tilde{\Phi} - \accentset{\circ}{\bf A}
\ee
where `$\accentset{\circ}{~}$' denotes the derivative with respect to $\tilde{\tau}$. Then, working again in the disformal Lorenz gauge: $\nabla \cdot {\bf A} = - \accentset{\circ}{\tilde{\Phi}} / c^2$, we arrive at the field-potential equations of motion:
\bs\label{eq:system}
\begin{eqnarray}
\left( \nabla^2 - \frac{1}{c^2}\frac{\partial^2}{\partial \tilde{\tau}^2} \right) \tilde{\Phi} 
&=& - \frac{\Omega}{\epsilon_0} \label{Phieq} \\
\left( \nabla^2 - \frac{1}{c^2}\frac{\partial^2}{\partial \tilde{\tau}^2} \right) {\bf A}  
&=& - \mu_0 \tilde{\bf J} \label{Aeq}.
\end{eqnarray}
\es
The system (\ref{eq:system}) is closed, and is now instantly recognizable from classical electrodynamics, hence easily solvable. Important to note here is that these tilde variables and coordinates just defined are exactly those we would have, had we originally expressed the action \ref{eq:action1} entirely in terms of the electromagnetic metric metric $g^{(r)}$ -- we are now working in the electromagnetic frame. 

%
\subsection{Field solutions and the bremsstrahlung condition}

The system of equations, \eqref{eq:system}, is readily satisfied by the Lienard-Wiechert potentials \cite{2007classical}. In terms of our electromagnetic frame quantities -- recalling that our coordinates are all comoving -- these solutions read 
\bs\label{eq:LW}
\ba
\tilde{\Phi}({\bf x}, \tilde{\tau}) &=&
\frac{Q}{4\pi \epsilon_0}~
\frac{1}{[1-{\bf n}(\tilde{\tau}') \cdot \boldsymbol\beta(\tilde{\tau}')]}~
\frac{1}{X(\tilde{\tau}')} \\
\mathbf{A}({\bf x},\tilde{\tau}) &=&
\mu_0 \epsilon_0{\bf V}(\tilde{\tau}')\tilde{\Phi}
\ea
\es
where $\tilde{\tau}'$ is the retarded electric frame time, defined by the implicit equation 
\be\label{eq:ret}
(\tilde{\tau}' - \tilde{\tau})c + X(\tilde{\tau}') = 0,
\ee
and we have made the following definitions 
\be\label{eq:defs}
X(\tau) := |{\bf x} - {\bf x}_p(\tau)|, \quad 
{\bf n}(\tau) := \frac{{\bf x} - {\bf x}_p(\tau)}{X(\tau)}, \quad
\boldsymbol\beta(\tau) := \frac{\tilde{\bf V}(\tau)}{c} = \frac{{\bf V}(\tau)}{c_s(\tau)}.
\ee
Combining \eqref{eq:Epot} with \eqref{eq:LW} and reversing the field redefinitions gives the following electric field profile in the Jordan frame:
\be\label{eq:profile}
\mathbf{E}({\bf x},\tau) = 
\frac{Q}{4\pi\epsilon(\tau)}
\left[
\frac{(1-\beta^2)({\bf n} - \boldsymbol\beta)}{ X^2[1 - {\bf n} \cdot {\boldsymbol \beta}]^3} + 
\frac{{\bf n} \times [({\bf n} - \boldsymbol\beta) \times \dot{\boldsymbol\beta}]}{c_sX[1 - {\bf n} \cdot {\boldsymbol \beta}]^3}
\right]_{\mathrm{ret}},
\ee
and, also in the Jordan frame:
\be
{\bf B}({\bf x},\tau) = \left[{\bf n}\times\frac{\bf E}{c_s}\right]_{\mathrm{ret}}
\ee
where we have reverted back to the $\tau$ time derivative, `$\dot{~}$', and quantities enclosed in the square brackets, $[...]_{\mathrm{ret}}$, are to be evaluated at the retarded time $\tau'$ given implicitly by equation \eqref{eq:ret} together with the relationship between $\tau$ and $\tilde\tau$ 
\be
\tau = \int\frac{d\tilde{\tau}}{Z(\tilde{\tau})}~,
\ee
which is non-local, and not analytically solvable in general. In the Jordan frame, the Poynting vector as obtained from Poynting's theorem is:
\be\label{eq:poyn_exp}
{\bf S} = {\bf E} \times \frac{{\bf B}}{\mu(\tau)}
\ee
and so, for a charged particle on a straight trajectory ($\boldsymbol\beta\times\dot{\boldsymbol\beta}=0$) the comoving power radiated, i.e. the power radiated per unit conformal time, is:
\be\label{eq:power}
{\mathcal P} = 
\frac{1}{4\pi\epsilon(\tau)}\frac{2Q^2}{3c_s(\tau)}\frac{1}{(1-\beta^2)^3}|\dot{\boldsymbol\beta}|^2,
\ee
obtained from \eqref{eq:poyn_exp}.

The second (radiative) term in equation \eqref{eq:profile} will be non-zero if and only if $|\dot{\boldsymbol\beta}|\neq0$. We can clearly see that, from the definition of $\boldsymbol\beta$ in equation \eqref{eq:defs}, this can be true even when the particle is not accelerating. If the comoving velocity ${\bf V}$ is constant, then $\dot{\boldsymbol\beta} = \boldsymbol\beta ~ \dot{c_s} / c_s$ and electromagnetic radiation will still carry energy outward, away from the particle\footnote{We note that the assumption of constant comoving velocity is non--trivial, due to the fact that we consider motion in an expanding background. Since we just want to consider the effect due to the disformal coupling, we ignore this issue here and refer the reader to \cite{Futamase:1996xh,Nomura:2006ka,Blaga:2014vna}.}. We sum this result up in the following radiation condition: a charged particle on an expanding background in motion -- uniform as seen by a stationary observer on the same background -- will in general radiate if the electromagnetic field couples to a second, distinct geometry, disformally varying with respect to the first: that is if $\dot{c}_s \neq0$.

We see in this setup that if the scalar field $\phi$ evolves in time (with $\dot Z$ non-zero), the particle appears to the electric field as one that is accelerating, even when in vacuum; to this phenomenon we attach the name vacuum bremsstrahlung. All this effect requires, really, is the condition that the speed of light vary with time. In fact, though we have demonstrated the  presence of vacuum bremsstrahlung in a disformally coupled field setting, it will no doubt be more widely applicable. We expect any theory in which the speed of light is dynamical in this sense to exhibit this phenomena, and hence to be testable in this way. We shall see in the next section, however, that for our theory, the effect is much smaller than Cherenkov radiation.

\subsection{Energy lost from a coupled cosmic ray}
\begin{figure}
\begin{center}
\includegraphics[width=\textwidth]{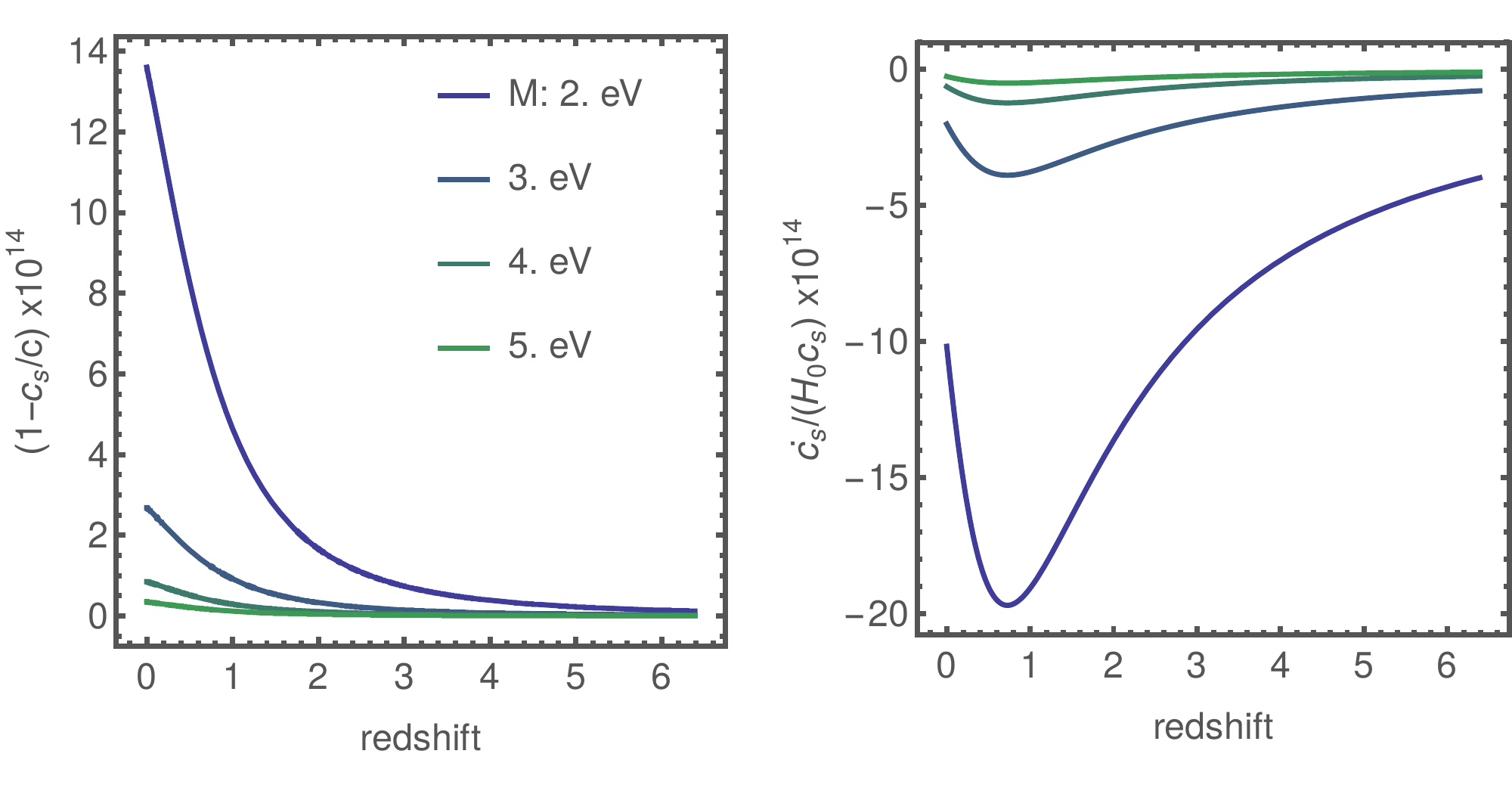}
\end{center}
\caption{Cosmological evolution of the observed speed of light, $c_s$, with redshift for values within LEP constraints (see Sec. \ref{sec:constraints}). $M$ is the energy scale associated to the disformal coupling -- defined in Eq. \eqref{eq:M} -- between light and a quintessence scalar field with exponential potential: $V(\phi)=V_0\mathrm{exp}[-\phi\kappa^{1/2}]$. \label{fig:Zs}}
\end{figure}
We will consider an ultra-high energy cosmic ray (a ray with energy in excess of about $10^{15}$ eV) in what follows. This means we can safely assume the cosmic ray travels along a straight line geodesic, that is $\boldsymbol\beta\times\dot{\boldsymbol\beta}=0$; intergalactic magnetic fields are extremely weak, too much so to curve the path of an ray of this energy appreciably. The radiation condition for expanding space is thus: vacuum bremsstrahlung will occur if $\dot{c}_s \neq 0$, even when the comoving velocity -- and hence the physical velocity -- is constant. In this case, we have that $\dot{\beta} = \beta \dot{Z} / Z$ and so both the square of the factor $\dot{Z} / Z$ and the sixth power of the Lorentz factor, $(1-\beta^2)^{-3}$, will determine the magnitude of energy lost by the particle through this process. 

If the scalar field $\phi$ is responsible for the late time accelerated expansion of the universe, then the cosmic ray's bremsstrahlung energy loss will be suppressed by the Hubble scale as measured in the present epoch (within a few redshift). Further, our model must obey the LEP constraints imposed on it in Sec. \ref{sec:constraints} which translates to an energy scale $M$ of roughly $\mbox{eV}$ or above. Both factors drive the allowed values of $\dot{Z} / Z$ down to very small values indeed for any viable cosmology scenario. We show in Fig. \ref{fig:Zs} several allowed evolution histories of the speed of light for a simple extension to the standard cosmological model, whereby the dark energy field is driven by an exponential potential with mild negative slope: $V(\phi)=V_0\mathrm{exp}[-\phi\kappa^{1/2}]$.

\begin{figure}
\begin{center}
\includegraphics[width=\textwidth]{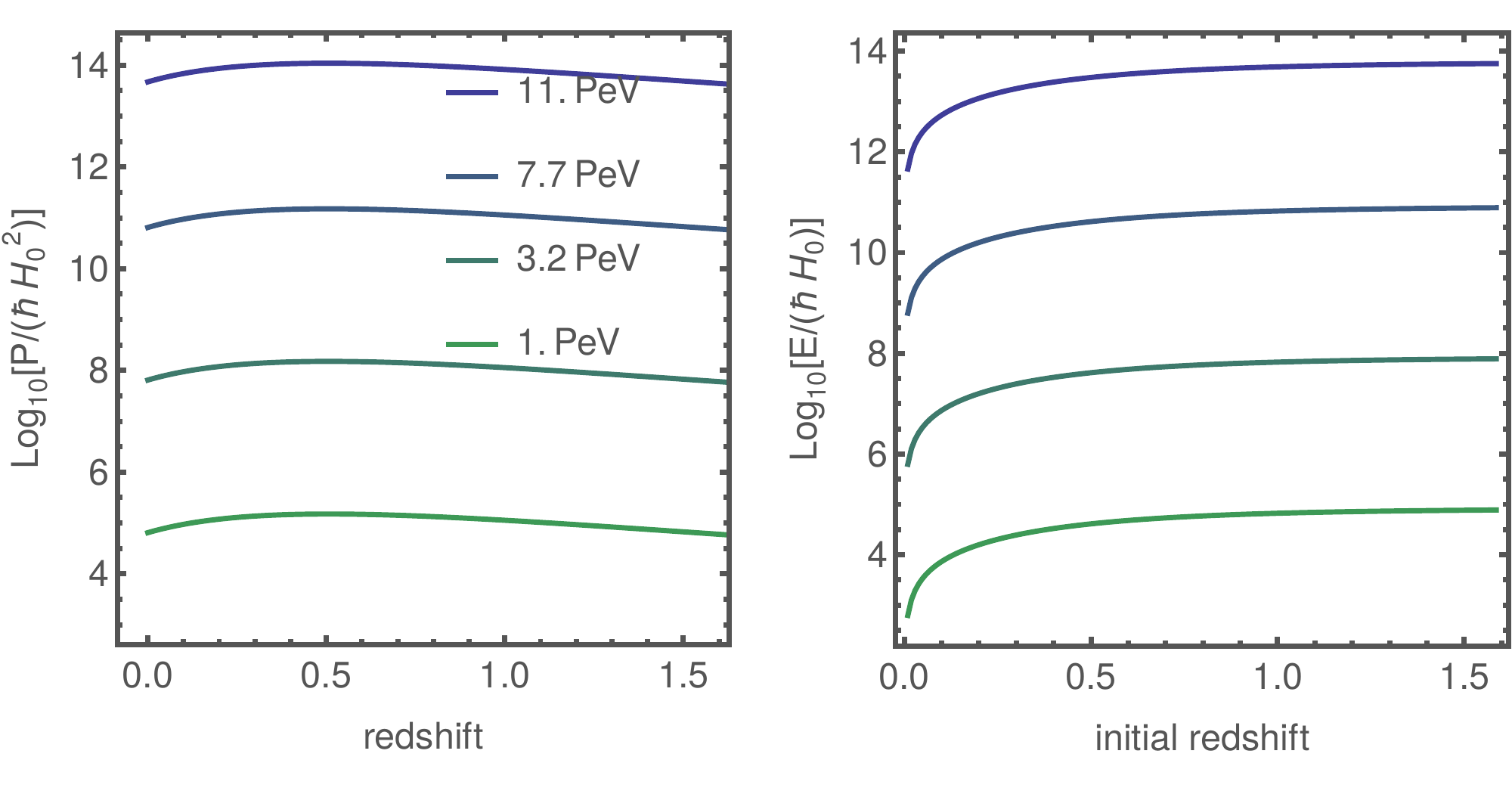}
\end{center}
\caption{\emph{Left:} Energy radiated per unit cosmic time $P={\mathcal P}a$ at each redshift, where $\mathcal{P}$ is defined in Eq. \eqref{eq:power}. \emph{Right:} Integrated energy loss by the particle for its entire trajectory, beginning at some initial redshift, arriving at earth today. In both plots, each curve corresponds to a cosmic ray with relativistic energy shown in the legend; PeV$=10^{15}$ eV. The disformal energy scale, $M$, is fixed at 2 eV for the left and right panels.\label{fig:PandE}}
\end{figure}

We see from the left panel of Fig. \ref{fig:Zs} that $\dot{c_s} / c_s$ must be very small -- many orders of magnitude \emph{less} than the Hubble scale at $H_0 \simeq 10^{-42} \mbox{ GeV}$! Observations of ultra high energy cosmic rays tell us we must consider charged particles with energy in excess of, say, a $\mbox{PeV}$, and bounds on $M$ from LEP mean that the expression for radiated power, Eq. (\ref{eq:power}) is valid up to very high velocity, but not above that for a few $\mbox{PeV}$, when vacuum Cherenkov radiation radically alters the nearby electric field behavior. In these cases the Lorentz factor is huge, and Eq. (\ref{eq:power}) shows the amount of power radiated by the ray is highly sensitive to the size of the Lorentz factor, yet, in Fig. \ref{fig:PandE} it is clear that this is not enough to beat the Hubble scale suppression. 

For most of these cosmic rays a galactic source is highly unlikely. More probable: they were accelerated by jets protruding from the active nuclei of quasars, some of which have been recorded by the Sloan Digital Sky Survey at cosmic distances of redshift up to about $z\simeq6$. However, the right panel of Fig. \ref{fig:PandE} shows that even the integrated energy loss across a distance this large is suppressed by the Hubble scale (as could perhaps be infered from dimensional analysis.) 
We conclude this section by remarking that such an effect will never be practically measurable if the disformal coupling is to dark energy. The Hubble scale today is so far from any of those in particle physics that a second order effect in a dynamic speed of light theory like vacuum bremsstrahlung will be negligible. For any such coupling or dynamic light speed in inflation is, however, a different story. The scale of inflation may be large enough that, during or just after reheating, these effects must be taken into account.

%
%
\section{Conclusions\label{sec:conc}}
We have shown that disformal couplings allow charged particles to emit Cherenkov radiation and bremsstrahlung in vacuum. The distortion of causal structure by the scalar field, a characteristic consequence of these interactions, can cause the speed of photons to be lower than that of a charged particle -- and even to vary in time -- mimicking a dielectric medium. 

To demonstrate this, we have developed a theory of electrodynamics in which a scalar field couples disformally to photons and charged particles, on both flat and expanding backgrounds. Unless the coupling strengths to each species are forced to be equal, two distinct frames appear in the theory, each with a specific role: working out observational quantities, such as the observed speed of light, required use of the frame in which matter is uncoupled from the scalar (i.e. the Jordan frame), but photons in general are not. Calculations were found to be simplest however, especially for a time dependent coupling, in the electromagnetic frame, where freely falling photons always follow geodesics.  

Working in flat space, we determined the constraints on dark energy models with disformal couplings that arise from the non-observation of vacuum Cherenkov radiation by the LEP collaboration. These parameter-space bounds are complementary to those obtained from spectral distortions of the CMB \cite{vandeBruck:2013yxa}; they both cover different regions and agree across their intersection. Finally, we have shown that the dark energy fine tuning problem is a problem for vacuum bremsstrahlung detection also: suppression of this particle physics interaction by the Hubble scale is unbeatable for any conceivable measurement one could dream of making on the earth's cosmic ray flux. 

In this study, we have converted the bounds on maximum attainable velocities of particles obtained by the LEP group to those on the scalar field coupling interaction $M$ via the Friedman equation. Explicitly, we: a) assumed our gravity sector was as simple as possible (quintessence with exponential potential, uncoupled to matter) and: b) produced constraints that are dependent on the measured dark energy equation of state today. This work should thus be extended along these two lines. How sensitive are these limits to changes in the gravitational sector of the theory? The bound on $M$, eq. (\ref{eq:constraintonM}), will change and this must be worked out. The interplay between particle physics and cosmology has so far been exceedingly rich, and constraining cosmological models, such as the present one, using results from ground-based particle experiments in this fashion remains a surprisingly fruitful venture. 

\vspace{0.5cm}\

\noindent {\bf Acknowledgements} We are grateful to J. Ronayne for a helpful comment. We also thank Robert Blaga for discussions on moving particles in an expanding spacetime. The work of CvdB is supported by the Lancaster-Manchester-Sheffield Consortium for Fundamental Physics under STFC Grant No. ST/L000520/1. CB is supported by a Royal Society University Research Fellowship.

%
%
\bibliography{bib}
\bibliographystyle{unsrt}

\end{document}